\documentclass[conference]{IEEEtran}
\IEEEoverridecommandlockouts
\usepackage{cite}
\usepackage[normalem]{ulem}
\usepackage{amsmath,amssymb,amsfonts}
\usepackage{algorithmic}
\usepackage{graphicx}
\usepackage{textcomp}
\usepackage{xcolor}
\usepackage{hhline}

\ifCLASSOPTIONcompsoc
    \usepackage[caption=false, font=normalsize, labelfont=sf, textfont=sf]{subfig}
\else
\usepackage[caption=false, font=footnotesize]{subfig}
\fi
\def\BibTeX{{\rm B\kern-.05em{\sc i\kern-.025em b}\kern-.08em
    T\kern-.1667em\lower.7ex\hbox{E}\kern-.125emX}}

\begin{document}

\title{No-Reference Video Quality Assessment Using Space-Time Chips}

\author{\IEEEauthorblockN{Joshua P. Ebenezer,\textsuperscript{\textsection}\footnotemark
\IEEEauthorrefmark{1} Zaixi Shang,\textsuperscript{\textsection}\IEEEauthorrefmark{1} Yongjun Wu,\IEEEauthorrefmark{2} Hai Wei,\IEEEauthorrefmark{2} Alan C. Bovik\IEEEauthorrefmark{1}}
\IEEEauthorblockA{\IEEEauthorrefmark{1} Laboratory for Image and Video Engineering (LIVE), The University of Texas at Austin \\
 \IEEEauthorrefmark{2} Amazon Prime Video \\
 joshuaebenezer@utexas.edu, zxshang@utexas.edu, yongjuw@amazon.com, haiwei@amazon.com, bovik@ece.utexas.edu}
}

\IEEEoverridecommandlockouts
\IEEEpubid{\makebox[\columnwidth]{978-1-4799-7492-4/15/\$31.00 \copyright2020
IEEE \hfill} \hspace{\columnsep}\makebox[\columnwidth]{ }}

\maketitle
\begingroup\renewcommand\thefootnote{\textsection}
\footnotetext{Equal contribution}
\endgroup

\begin{abstract}
We propose a new prototype model for no-reference video quality assessment (VQA) based on the natural statistics of \textit{space-time chips} of videos. Space-time chips (ST-chips) are a new, quality-aware feature space which we define as space-time localized cuts of video data in directions that are determined by the local motion flow. We use parametrized distribution fits to the bandpass histograms of space-time chips to characterize quality, and show that the parameters from these models are affected by distortion and can hence be used to objectively predict the quality of videos. Our prototype method, which we call ChipQA-0, is agnostic to the types of distortion affecting the video, and is based on identifying and quantifying deviations from the expected statistics of natural, undistorted ST-chips in order to predict video quality. We train and test our resulting model on several large VQA databases and show that our model achieves high correlation against human judgments of video quality and is competitive with state-of-the-art models. 
\end{abstract}


\section{Introduction}
Video content accounts for a very large portion of traffic on the Internet and continues to surge in volume, as more people stream content on smartphones, tablets, and high-definition screens. Being able to predict  perceived video quality is important to content providers for monitoring and controlling streaming quality and thereby enhance customer satisfaction. Video quality is affected when the video is distorted, which occurs due to a number of reasons as the video is being captured, transmitted, received, and displayed. The task of predicting the quality of a distorted video without a pristine video of the same content to compare it with, which is called no-reference (NR) VQA, is difficult. NR VQA is a hard problem because less information is available than models that use a reference, and conventional notions of signal fidelity are not applicable. Here we describe a new NR VQA algorithm based on the statistics of local windows oriented in space and time. 

 We briefly review state-of-the-art NR VQA algorithms. V-BLIINDS\cite{vbliinds} is a distortion agnostic NR VQA algorithm that models the statistics of DCTs of frame differences to predict video quality. These features are based on models of the human visual system (HVS). HVS-based algorithms posit that natural video statistics have a regularity from which the statistics of distorted videos deviate, and which the HVS is attuned to. Human perceptual judgments of quality are influenced by the degree to which the statistics of distorted videos deviate from those of natural videos. Such models thus attempt to understand and mimic operations of the HVS in order to identify the regularity in natural video statistics. VIIDEO\cite{viideo} is another algorithm that is belongs to this category. VIIDEO is completely "blind," in the sense that it does not require any training and can be directly deployed. VIIDEO makes use of observed high inter subband correlations of natural videos to predict quality. TLVQM\cite{tlvqm} takes a different approach to the problem, where the goal is to define features that capture distortion, and not to characterize naturalness per se. A number of spatiotemporal features are defined  at two computational levels, collectively designed to capture a wide range of distortions. TLVQM has about 30 different user-defined parameters, which may affect its performance on databases it was not exposed to when it was designed. 

It has also been observed \cite{zhengzhong} that NR image quality assessment algorithms work reasonably well when applied frame-by-frame to distorted videos of user-generated content because of a lack of temporal variation in such videos. FRIQUEE\cite{friquee} is a state-of-the-art algorithm for NR IQA that uses a bag of perceptually motivated features. BRISQUE\cite{brisque} is another NR IQA algorithm that models the statistics of spatially bandpassed coefficients of images, motivated by the fact that the early stages of the HVS perform spatial bandpassing. NIQE\cite{niqe} also models statistics of spatially bandpassed coefficients, but generates an opinion score by quantifying the deviation of a distorted image from the statistical fit to a corpus of natural images and does not require training. CORNIA\cite{cornia} approaches the problem differently from HVS-based models, and builds a dictionary to represent images effectively for quality assessment.

 Videos are spatiotemporal signals and distortions affecting videos can be spatial or temporal or a combination of both. An effective NR VQA algorithm must be able to build a robust representation of the spatiotemporal information in videos. Primary visual cortex (area V1) is implicated in decomposing visual signals into orientation and scale-tuned spatial and temporal channels. V1 neurons are sensitive to specific local orientations of motion. This decomposition is passed on to other areas of the brain, including area middle temporal (MT), where further motion processing occurs. Extra-striate cortical area MT is known to contain neurons that are sensitive to motion over larger spatial fields\cite{movie,simoncelli}. We propose a perceptually motivated NR VQA model, ChipQA, that captures both spatial and temporal distortions, by building a representation of local spatiotemporal data that is  attuned to local orientations of motion but is studied over large spatial fields. We show how observed statistical regularities of spatially bandpassed coefficients can be extended temporally and introduce the notion of space-time chips, which follow natural video statistics (NVS). We evaluate the new model on a number of databases and show that we are able to achieve state-of-the-art performance at reasonable computational cost.
\section{Proposed algorithm}

\subsection{Space-time Chips}
If we consider videos as space-time volumes of data, Space-time (ST) chips can be defined as chips of this volume in any direction or orientation. ST-Chips are similar to space-time slices\cite{sts1,sts2,sts3}, but are highly localized in space and time. We are interested in finding the specific directions and orientations of these chips that capture the regularity of natural videos. Consider a frame $I_T$ at time instant $T$, and its preceding frames $I_{T-T'+1}..I_{T-1}$ from time instant $T-T'+1$ onward, all of size $M \times N$. There are $T'$ frames in this space-time volume. Assume that we have coherent motion vectors available to us at time instant $T$. To the best of our knowledge, localized groups of pixels are in motion along the directions of these vectors. Assuming $T'$ is small enough, a local spatiotemporal chip of this volume that is oriented perpendicular to the motion vector at each spatial location would capture the local areas that are in motion at each location and across time. 

Taking into account the smoothness of motion across time for natural videos, we expect ST-chips to follow similar regularities as we would expect to find for stationary frames. Spatially bandpassed coefficients are known to follow a generalized Gaussian distribution (GGD) in the first order, and their second order statistics can be approximated as following an asymmetric generalized Gaussian distribution (AGGD)\cite{niqe,brisque}. Accordingly, we define the mean subtracted contrast normalized (MSCN) coefficients
\begin{equation}
    \hat{I}_T(i,j) = \frac{I_T(i,j)-\mu_T(i,j)}{\sigma_T(i,j)+C}
 \end{equation}
  where $(i,j)$ are spatial coordinates and we define the local mean and local variance as
  \begin{equation}
  \mu_T(i,j) = \sum\limits_{k=-K}^{k=K} \sum\limits_{k=-L}^{k=L} w(k,l) I_T(i+k,j+l)
  \end{equation}
\begin{equation}
\sigma_T(i,j) = \sqrt{\sum\limits_{k=-K}^{k=K} \sum\limits_{k=-L}^{k=L} w(k,l) (I_T(i+k,j+l)-\mu_T(i,j))^2}
\end{equation}
respectively. $w = \{w(k,l), k\in -K,..,K, l\in -L,..,L\}$ is a 2D circularly-symmetric Gaussian weighting function sampled out to 3 standard deviations and rescaled to unit volume.

\begin{figure}
  \includegraphics[width=\linewidth]{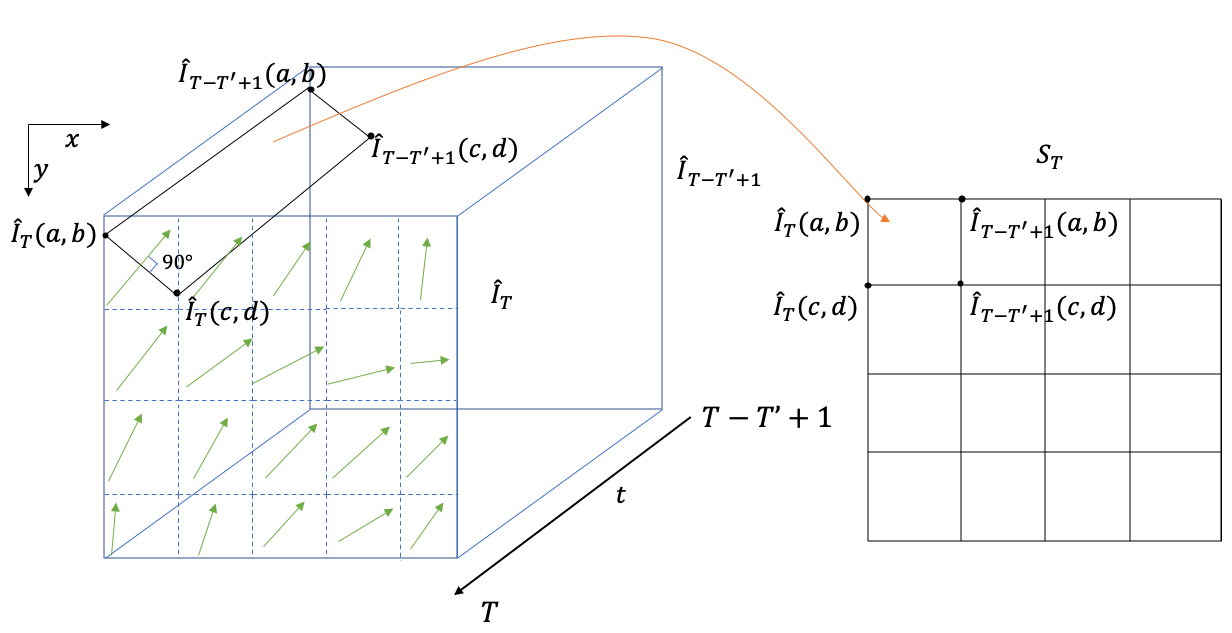}
  \caption{Extracting ST-chips. On the left is a spatiotemporal volume of frames from time $T-T'+1$ to $T$. The green arrows represent motion vectors at each spatial patch at time $T$. ST-chips are cut perpendicular to these across time and aggregated across spatial patches to form the frame on the right at each time instance.}\label{fig:stchip}
\end{figure}

Motion has been used to guide video quality prediction in several full-reference models\cite{movie,motionstruct,frof}, but few NR models\cite{vbliinds,nrof}. Given a dense motion flow field at time $T$, we first find the median of the flow field across a spatial window of size $R \times R$. This gives us a more robust estimate of  the flow in each spatial window. We cut the video volume of MSCN coefficients across time from $\hat{I}_{T-T'+1}$ to $\hat{I}_T$ at each window in the direction perpendicular to the motion vectors at each spatially localized window, as shown in Fig. \ref{fig:stchip}, and then aggregate these chips across the spatial windows to form a single ``frame" $S_T$ of ST-Chips at each time instance. Each chip is constrained to pass through the center of the spatial $R\times R$ patch, and to be perpendicular to the median flow of that patch. $R$ pairs of $x$ and $y$ coordinates obtained from the relevant line equations are rounded to integers for sampling from the video.In this way, the chips at each patch are uniquely defined by the spatial location of the patch and the direction of the median motion vector. These frames do not have well defined axes because each ST chip is oriented differently in space-time, but they contain important spatiotemporal data. For simplicity, we define $R=T'$. $S_T$ then has dimensions $ M' \times N'$, where $M'=T'\lfloor \frac{M}{T'} \rfloor$ and $N'= T' \lfloor \frac{N}{T'}\rfloor$.

\begin{figure} 
  \includegraphics[width=\linewidth]{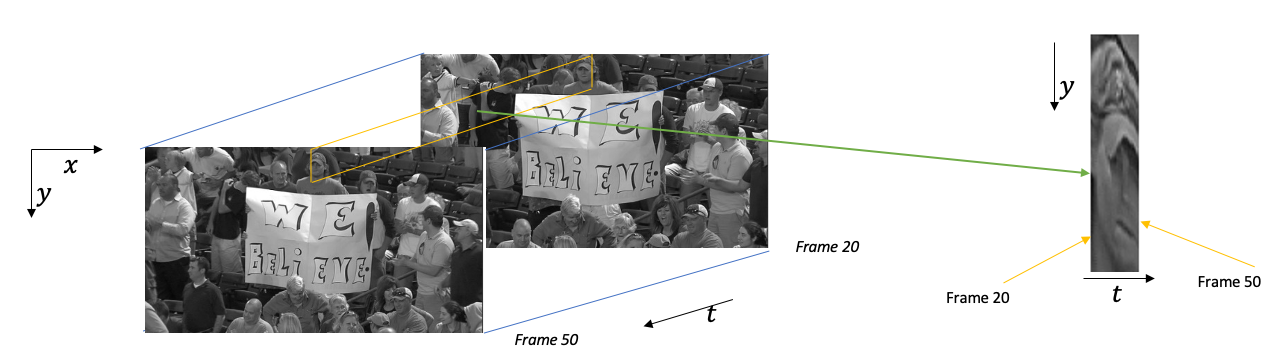}
  \caption{ST-Chips capture views of objects in motion. In this video, the person in the center moves to their right over time. Consequently, a chip taken in the proximity of their face over this duration and perpendicular to their motion captures their face itself.}\label{fig:proof}
\end{figure}

Ideally, the directions we have defined are the ones most likely to capture objects in motion. To see this, consider an object in motion in a fixed direction. An ST chip that is perpendicular to the direction of motion of the object is a plane that cuts through the cross-section of the object as it moves along time. A concrete example of this is given in Fig. \ref{fig:proof} for a video before its MSCN coefficients are computed.

\begin{figure*}[ht!]
\centering
\subfloat[Aliased and pristine]{{\includegraphics[width=0.18\textwidth]{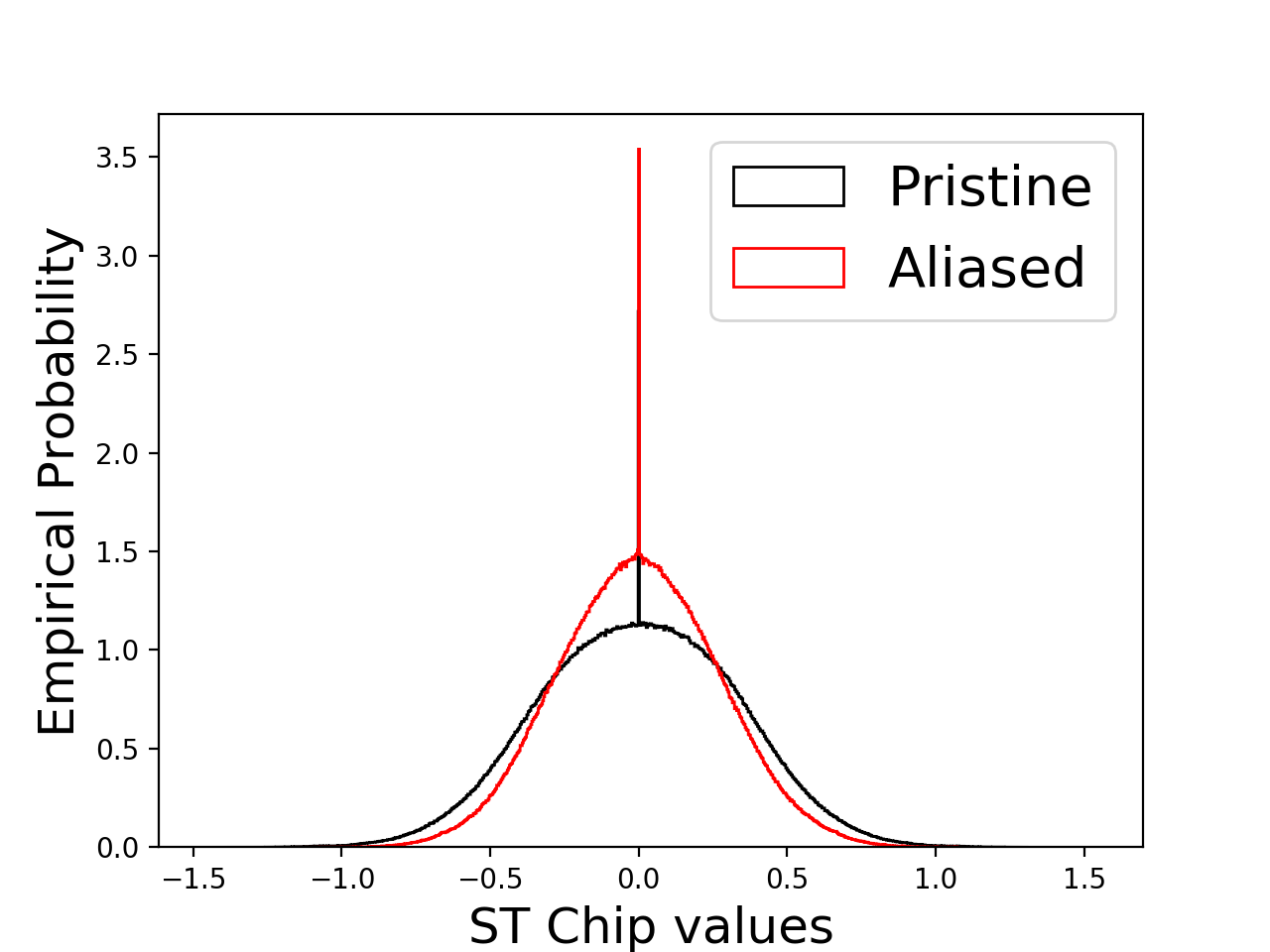}}}
\subfloat[Judder and pristine]{{\includegraphics[width=0.18\textwidth]{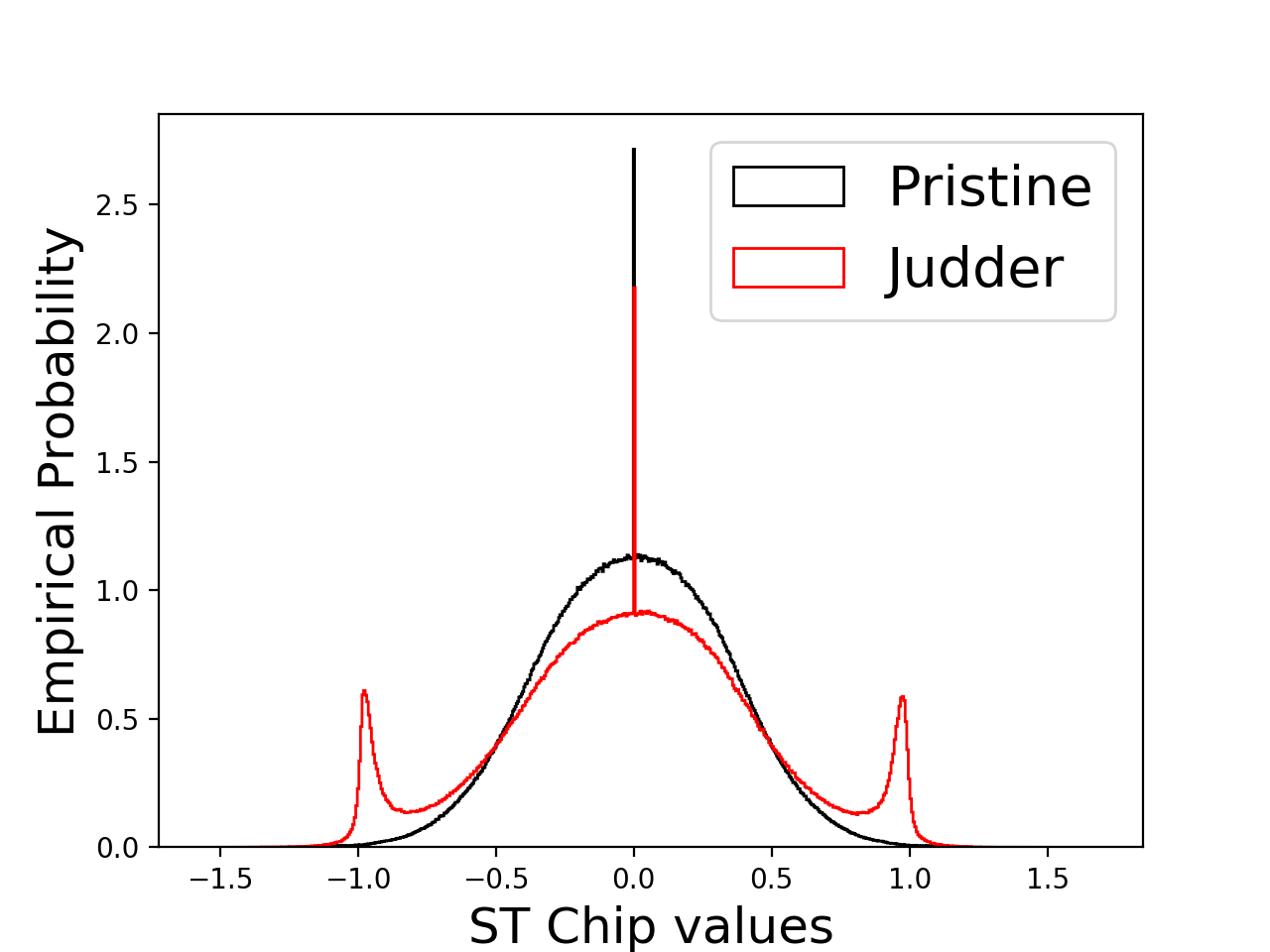}}} 
\subfloat[Flicker and pristine]{{\includegraphics[width=0.18\textwidth]{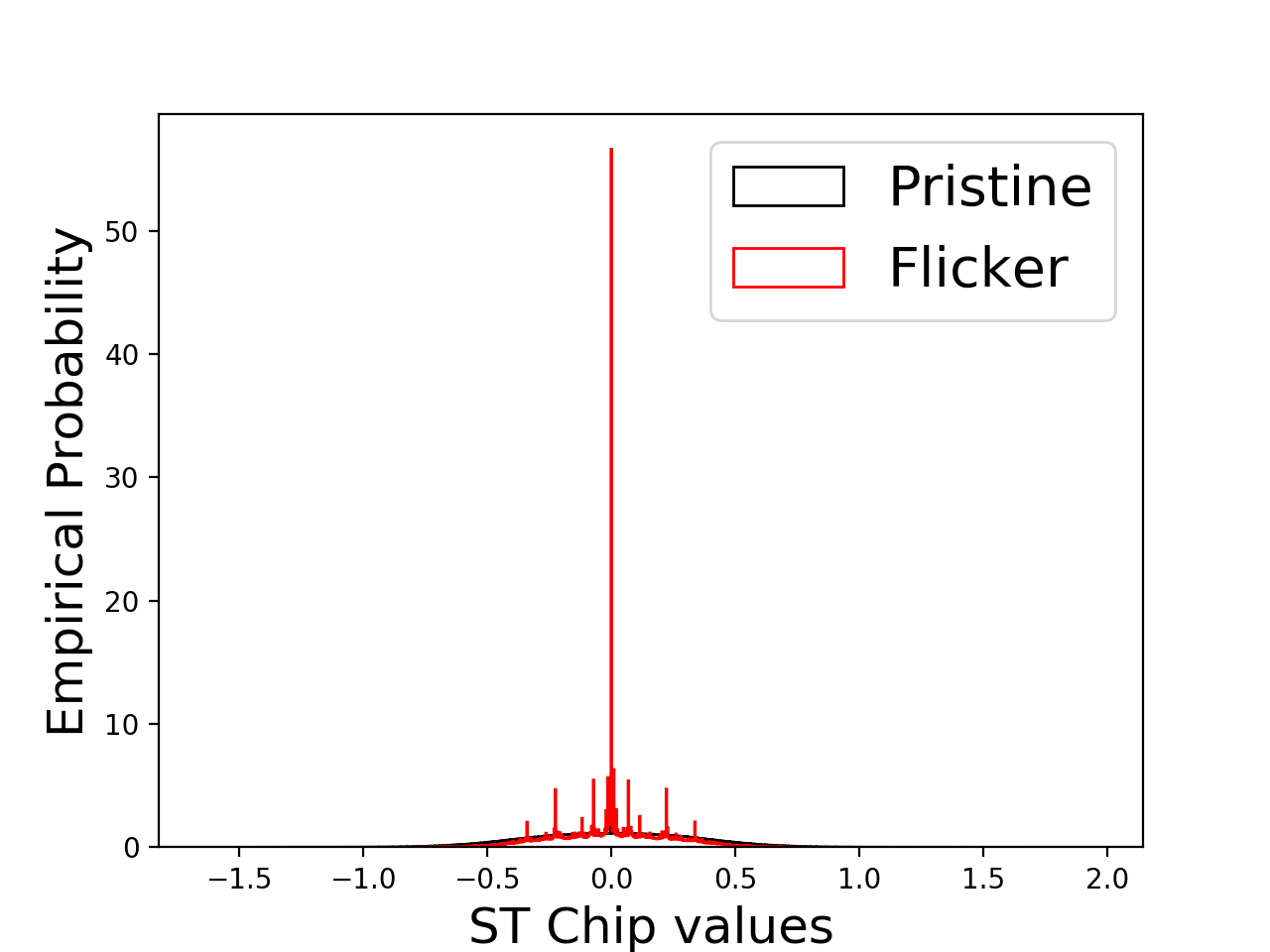}}}
\subfloat[Interlaced and pristine]{{\includegraphics[width=0.18\textwidth]{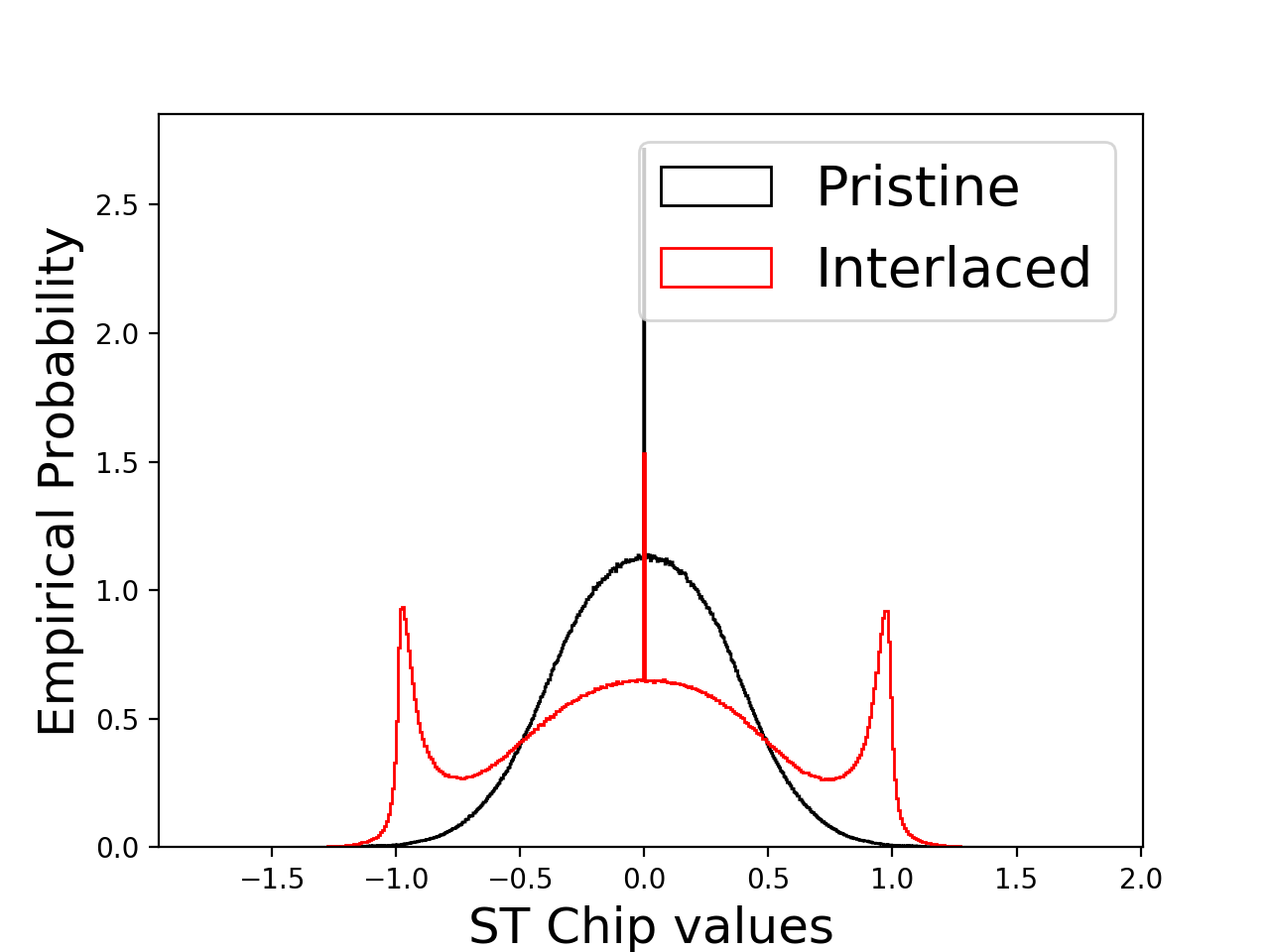}}}
\caption{Empirical distributions of ST-Chips. Pristine (original) distributions are in black and distorted distributions are in red.  }
\label{fig:stchips}
\end{figure*}

\begin{figure*}[ht]
\centering
\subfloat[Compressed and pristine]{{\includegraphics[width=0.18\textwidth]{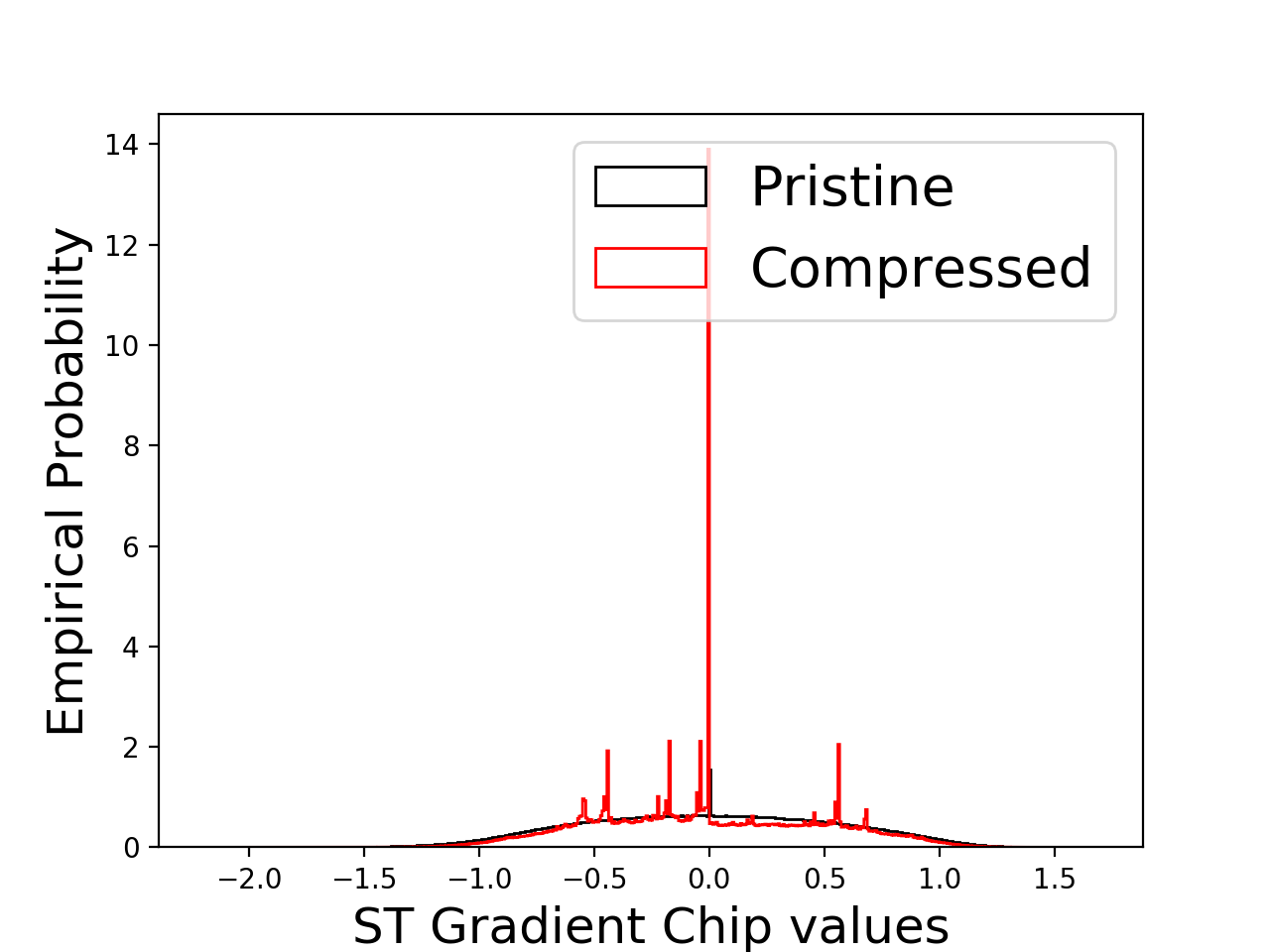}}}
\subfloat[Frame Drop and pristine]{{\includegraphics[width=0.18\textwidth]{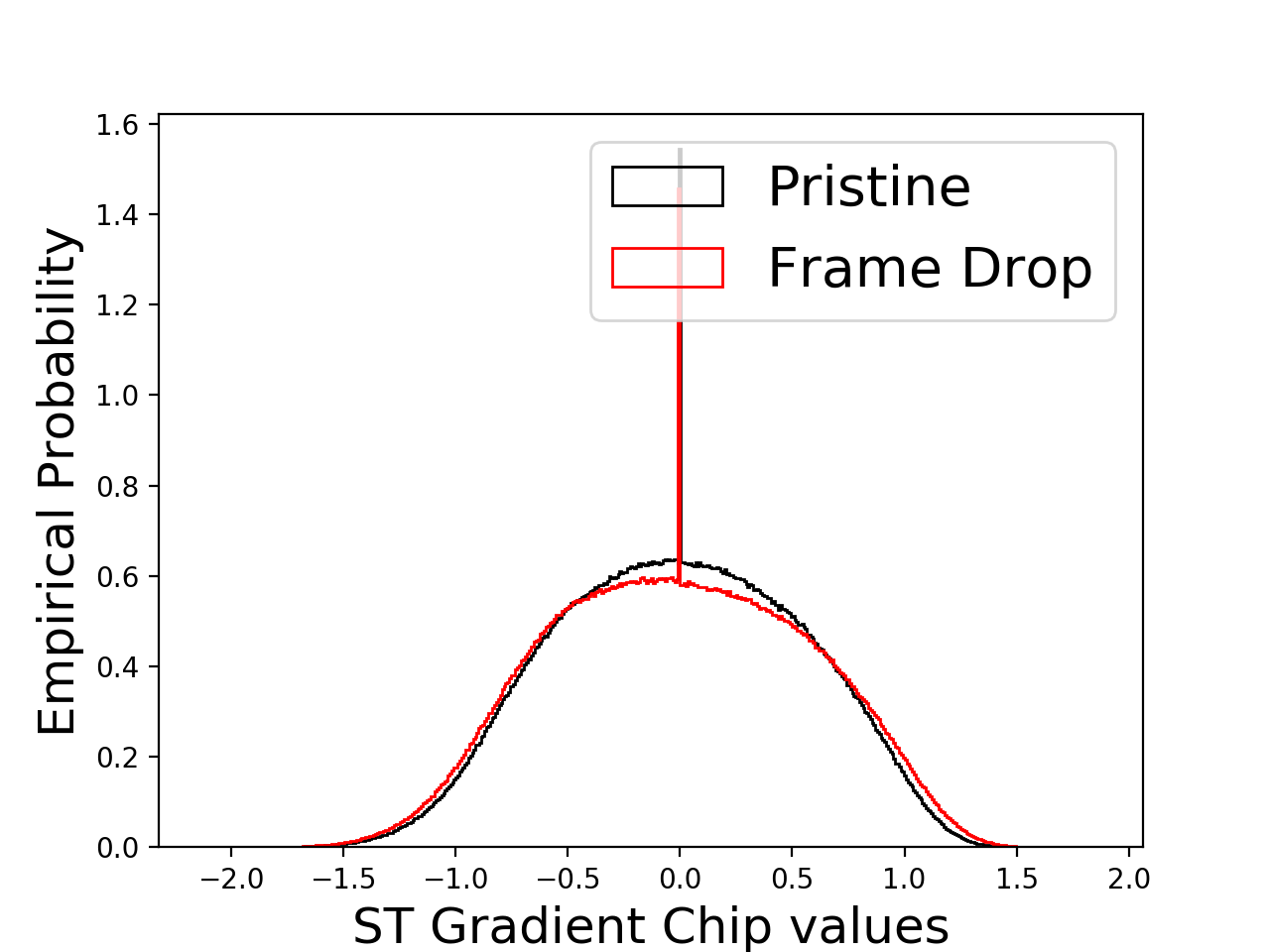}}} 
\subfloat[Flicker and pristine]{{\includegraphics[width=0.18\textwidth]{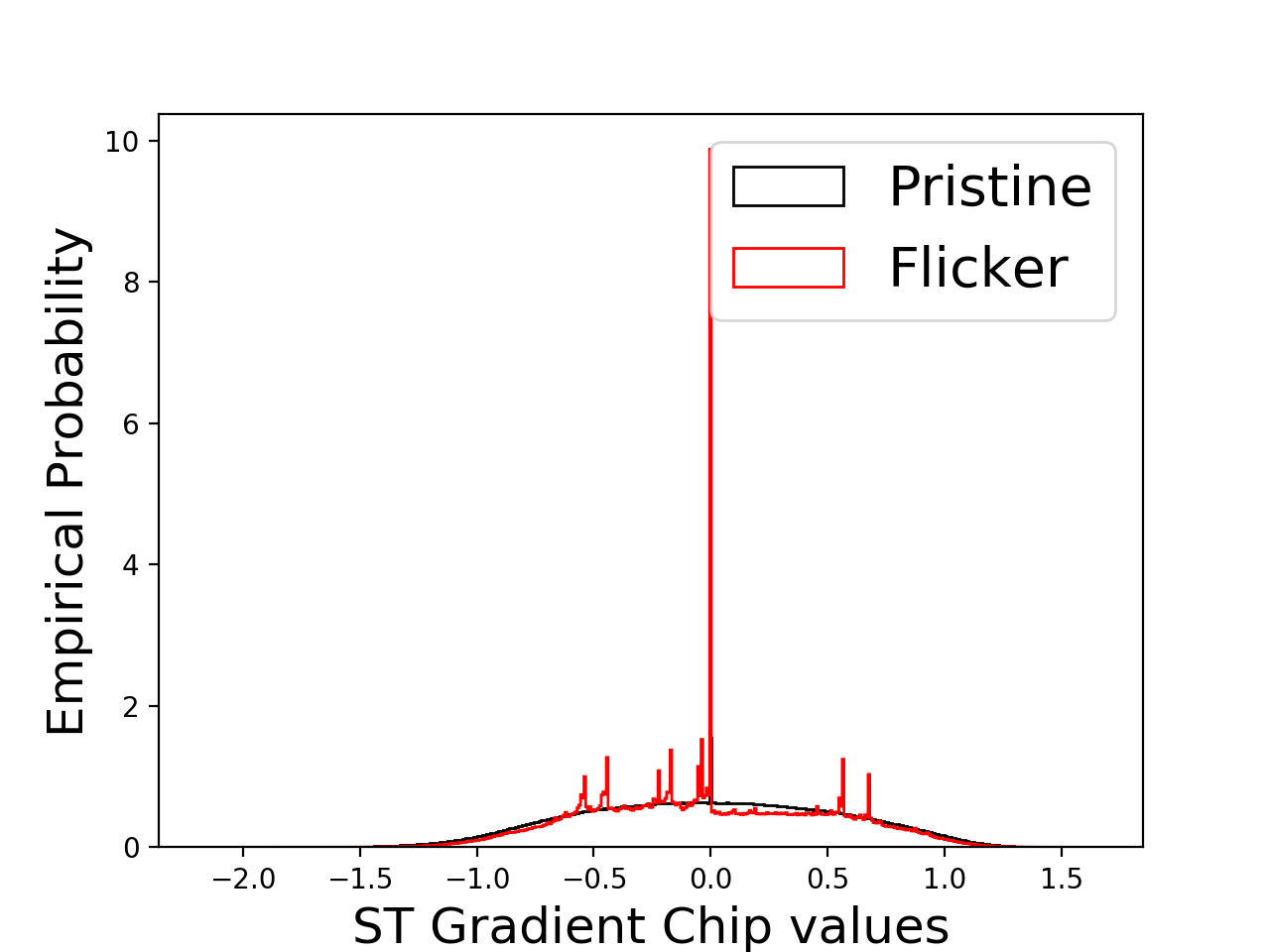}}}
\subfloat[Judder and pristine.]{{\includegraphics[width=0.18\textwidth]{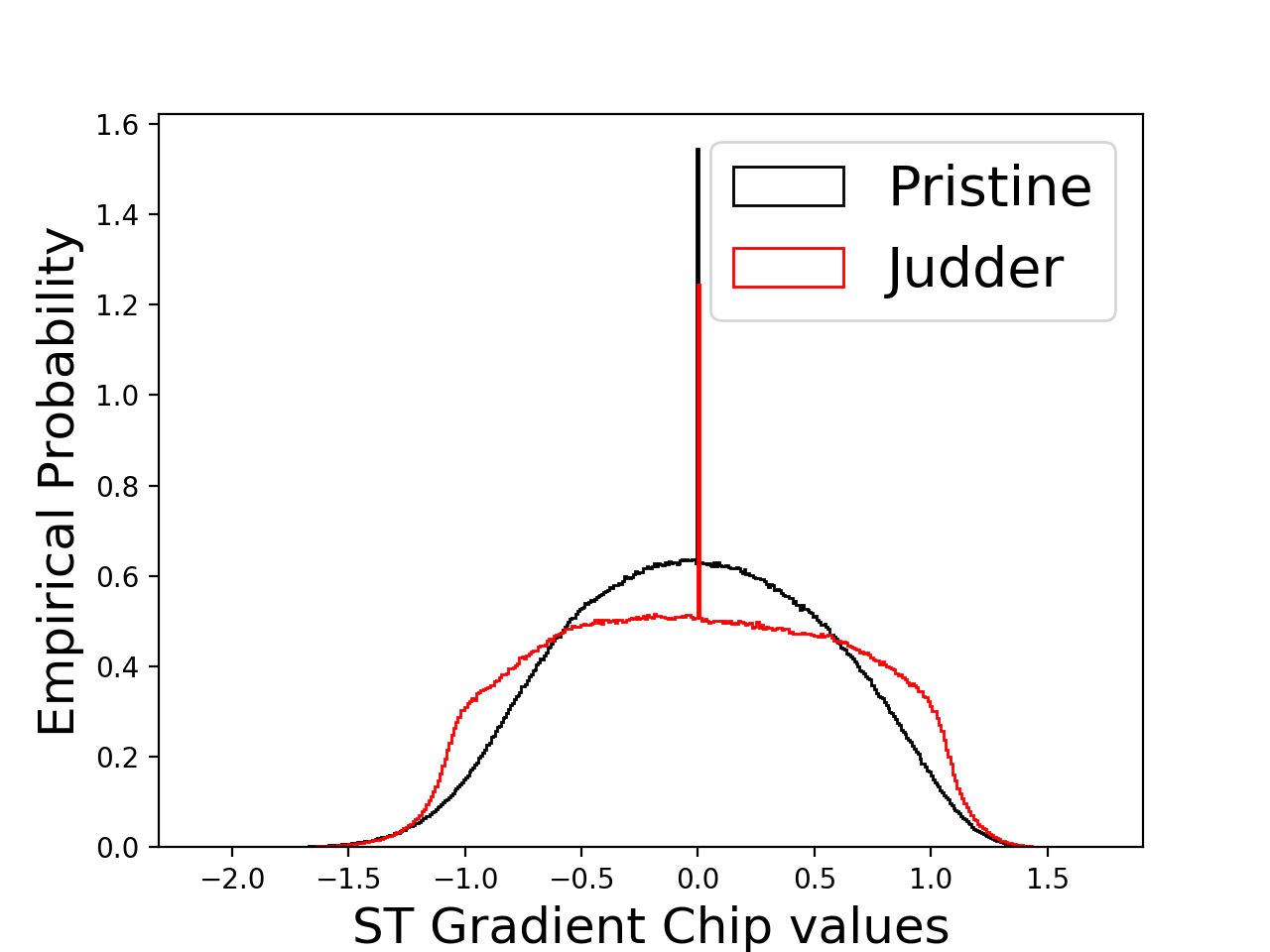}}}
\caption[]{Empirical distributions of ST Gradient chips. Pristine (original) distributions are in black and distorted distributions are in red.}
\label{fig:stgradchips}
\end{figure*}

\begin{figure*}[ht]
\centering
\subfloat[Aliased and pristine]{{\includegraphics[width=0.18\textwidth]{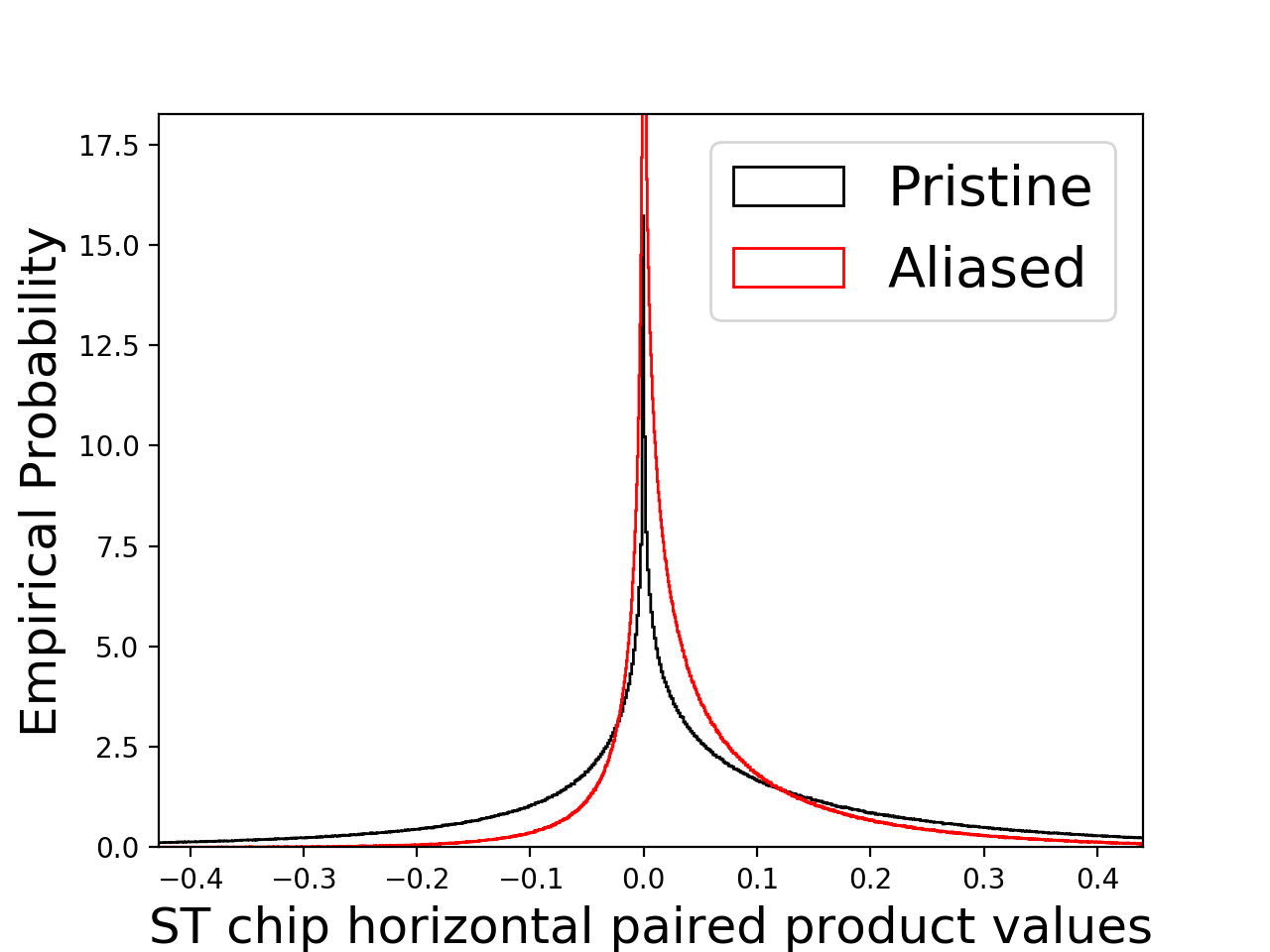}}}
\subfloat[Interlaced and pristine]{{\includegraphics[width=0.18\textwidth]{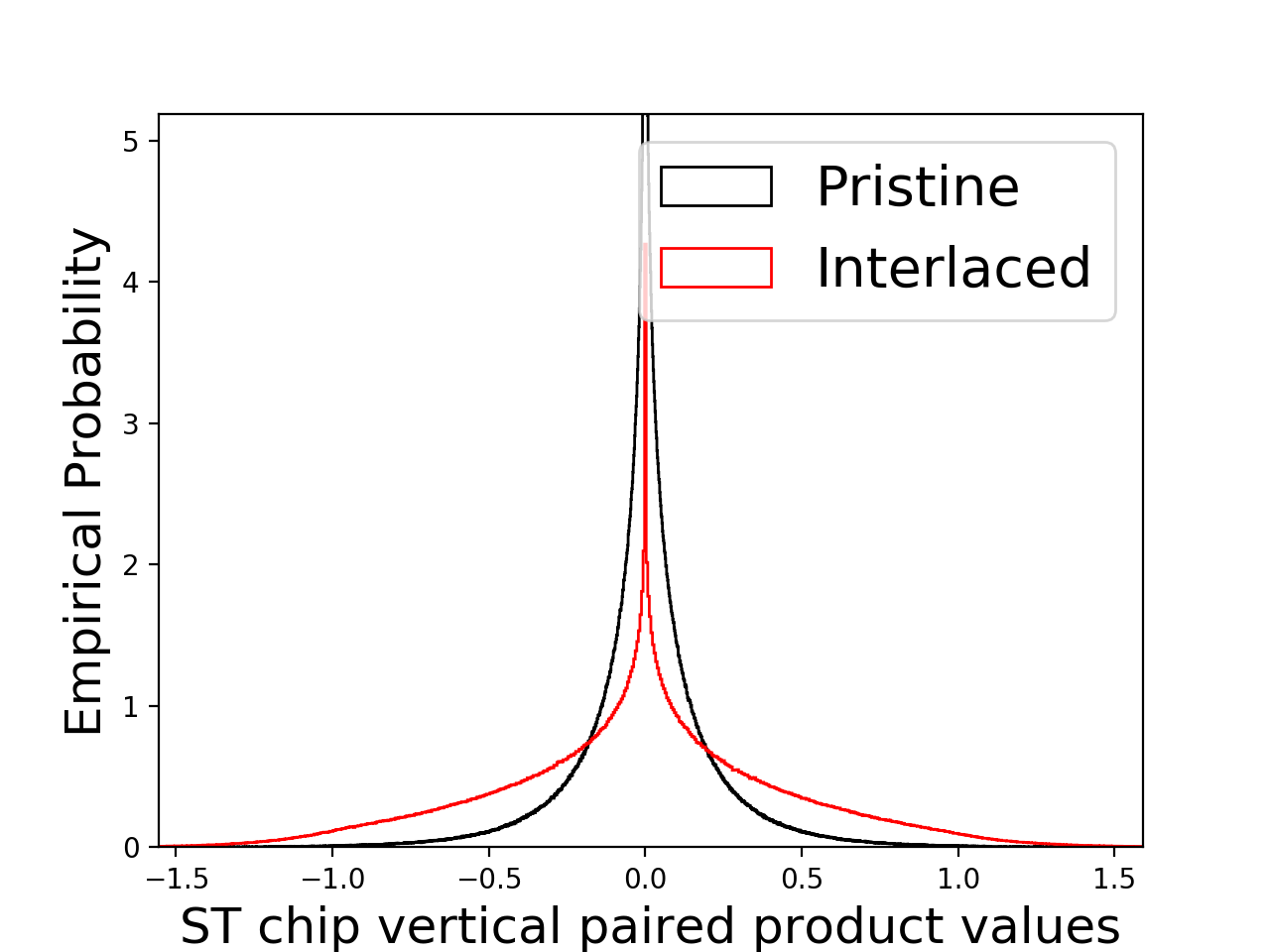}}} 
\subfloat[Compressed and pristine]{{\includegraphics[width=0.18\textwidth]{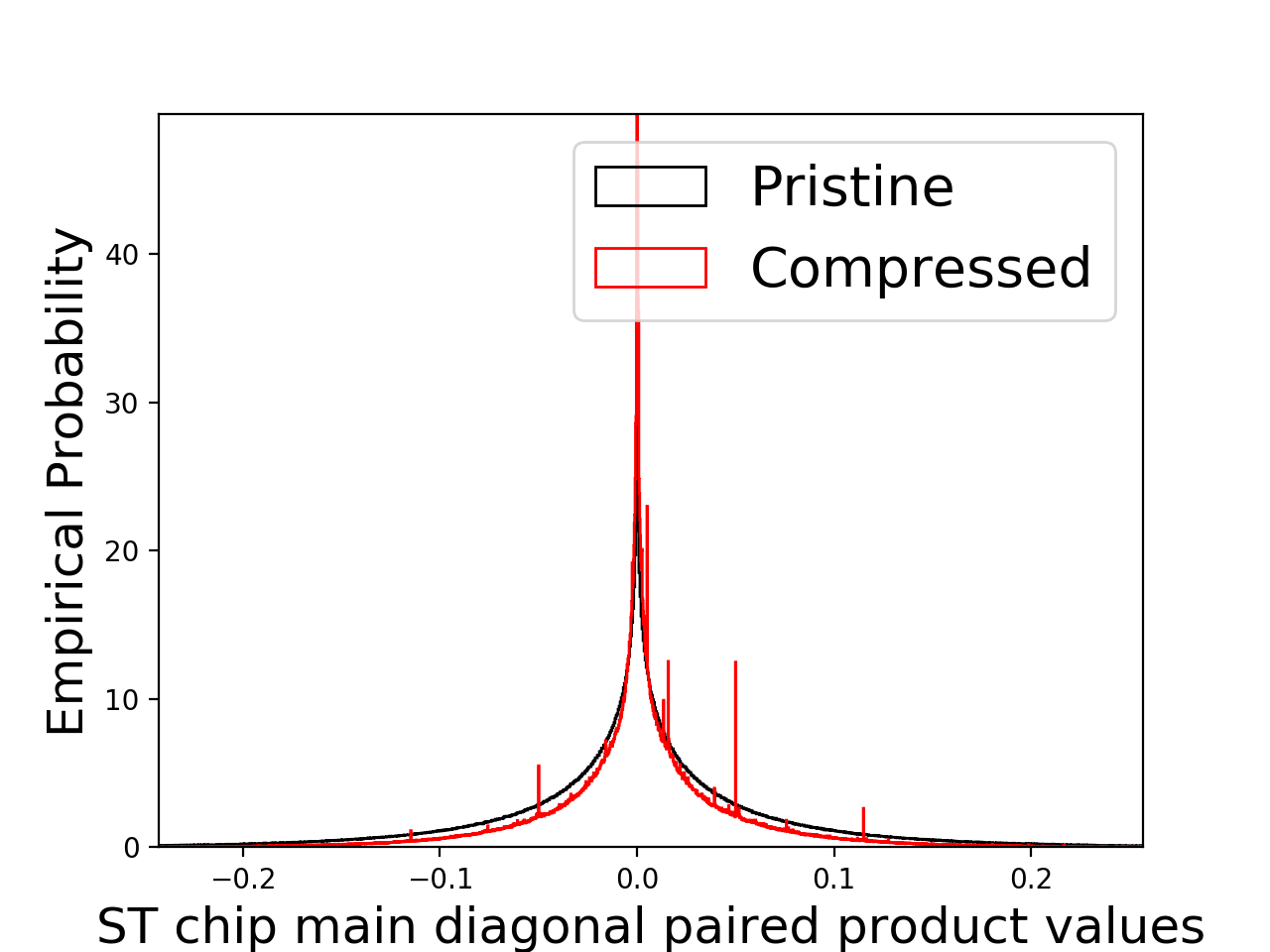}}}
\subfloat[Frame drop and pristine]{{\includegraphics[width=0.18\textwidth]{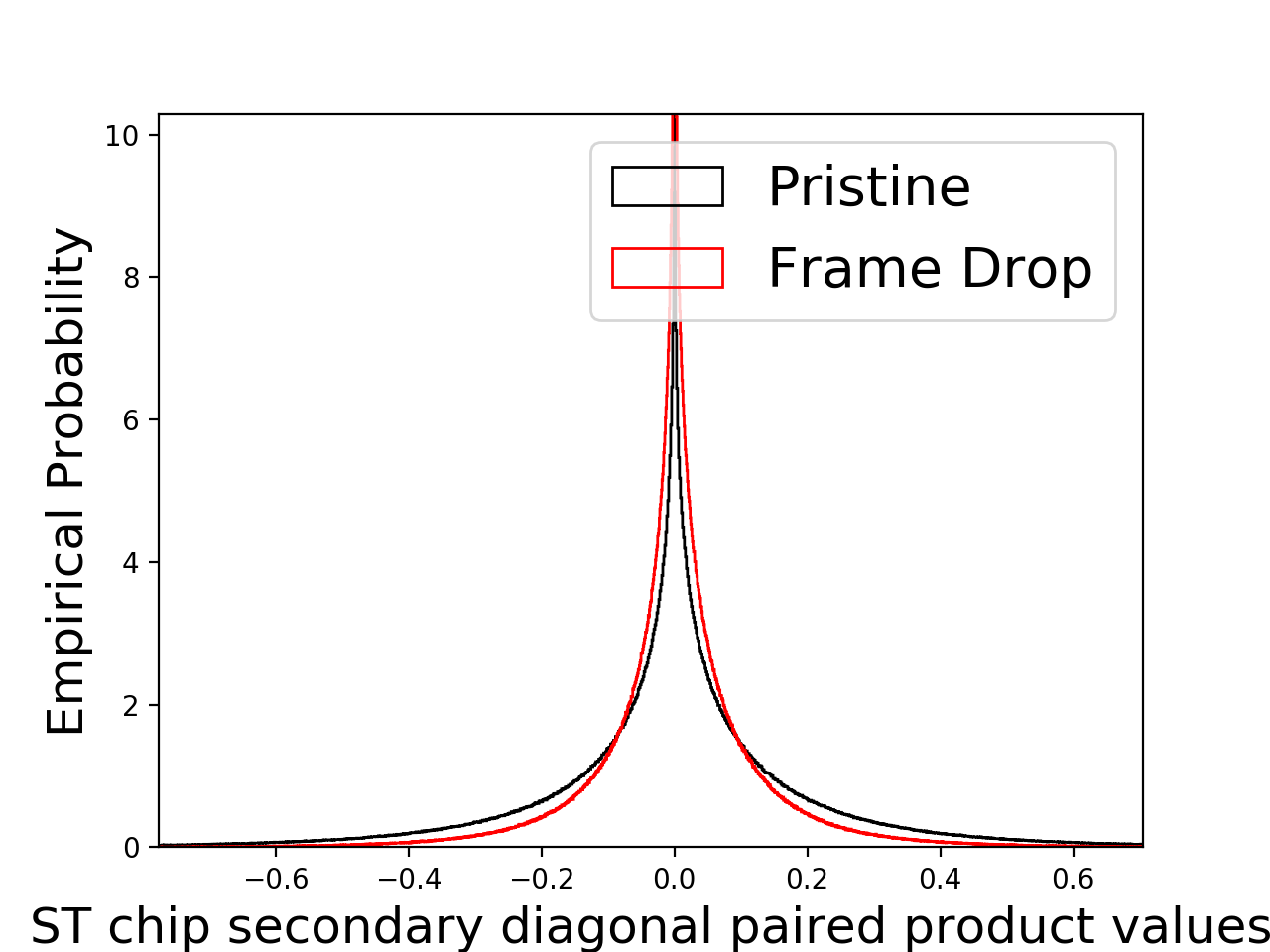}}}
\caption{Empirical distributions of paired products of ST-Chips. Pristine (original) distributions are in black and distorted distributions are in red.}
\label{fig:stpairedchips}
\end{figure*}

  ST-Chips are collected over all spatial patches of MSCN coefficients for every group of preceding $T'$ frames at each time instant $T$ and are then aggregated. We use the Farneback\cite{farneback} optical flow algorithm in all our experiments. Due to the smoothness of motion over time and the fact that we expect these chips to have captured natural objects in motion, we hypothesize that this aggregated spatiotemporal data will follow a regular form of natural video statistics. Spatiotemporal distortions are expected to affect the optical flow as well as spatial data. The motion estimates from optical flow can be distorted for distorted videos, and hence we would expect to see a deviation from the statistics of chips computed from pristine videos. We would also expect to see a deviation in the statistics of the MSCN coefficients collected across time, as distortions affect their regularity across time and across space. Both the distortion of MSCN coefficients and the distortion of optical flow are expected to contribute to a deviation from the statistics of a pristine video, and we confirm this empirically, as shown in Figs. \ref{fig:stchips}, \ref{fig:stgradchips}, and \ref{fig:stpairedchips}. ST-Chips of MSCN coefficients are found to follow a generalized Gaussian distribution (GGD) of the form:
  \begin{equation}
  f(x;\alpha;\beta) = \frac{\alpha}{2\beta \Gamma(\frac{1}{\alpha})} \exp (-(\frac{|x|}{\beta})^\alpha)
  \end{equation}
where $\Gamma(.)$ is the gamma function:
\begin{equation}
\Gamma(\alpha) = \int_0^\infty t^{\alpha-1} \exp(-t) dt.
\end{equation}
The shape parameter $\alpha$ of the GGD and the variance of the distribution are estimated using the moment-matching method described in \cite{estimation}.

We also model the second-order statistics of ST-Chips.  Define the collection of ST-Chips aggregated at each time instance $T$ as $S_T$, and define the pairwise products
\begin{equation}
H_T(i,j) = S_T(i,j)S_T(i,j+1)
\end{equation}
\begin{equation}
V_T(i,j) = S_T(i,j)S_T(i+1,j)
\end{equation}
\begin{equation}
D1_T(i,j) = S_T(i,j)S_T(i+1,j+1)
\end{equation}
\begin{equation}
D2_T(i,j) = S_T(i,j)S_T(i+1,j-1)
\end{equation}
These pairwise products of neighboring ST chip values along four orientations are modeled as following an asymmetric generalized Gaussian distribution (AGGD), which is given by:
\begin{equation}
f(x;\nu,\sigma_l^2,\sigma_r^2) = \begin{cases}
\frac{\nu}{(\beta_l+\beta_r)\Gamma (\frac{1}{\nu})} \exp(-(-\frac{x}{\beta_l})^\nu) &  x<0 \\
\frac{\nu}{(\beta_l+\beta_r)\Gamma (\frac{1}{\nu})} \exp(-(\frac{x}{\beta_r})^\nu) &  x>0 
\end{cases} 
\end{equation}
where
\begin{equation}
\beta_l = \sigma_l \sqrt{\frac{\Gamma(\frac{1}{\nu})}{\Gamma(\frac{3}{\nu})}}
\end{equation}
\begin{equation}
\beta_r = \sigma_r \sqrt{\frac{\Gamma(\frac{1}{\nu})}{\Gamma(\frac{3}{\nu})}}
\end{equation}
$\nu$ controls the shape of the distribution and $\sigma_l$ and $\sigma_r$ control the spread on each side of the mode. The parameters ($\eta,\nu,\sigma_l^2,\sigma_r^2$) are extracted from the best AGGD fit to each pairwise product, where 
\begin{equation}
\eta = (\beta_r-\beta_l) \frac{\Gamma(\frac{2}{\nu})}{\Gamma(\frac{1}{\nu})}.
\end{equation}
Videos are affected at multiple scales by distortions, and so all the features defined above are extracted at a reduced resolution as well. Each frame is low-pass filtered and downsampled by a factor of 2. Motion vectors are computed at the reduced scale and ST-Chips are extracted, as described previously, from volumes of MSCN coefficients at the reduced scale.

\subsection{Spatiotemporal Gradient Chips}

The spatial gradients of videos contain important information about edges and corners. Distortions are often more noticeable around edges and affect gradient fields. For this reason, we compute the magnitude of the gradient  for each frame using the Sobel filter. We then extract ST-Chips from spatiotemporal volumes of the MSCN coefficients of the gradient magnitude at two scales. The first-order and second-order statistics are modeled as GGD and AGGD, respectively, as described earlier for MSCN coefficients of pixel data. 

\subsection{Spatial features}

Spatial features and a naturalness score are computed frame-by-frame with the image naturalness index NIQE. These capture purely spatial aspects of distortion that may not be completely contained in ST-Chips, and hence boost performance. 

\subsection{Quality assessment}

Table~\ref{feat} gives a summary of all the features used in the prototype algorithm, ChipQA-0. A total of 109 features are extracted at each time instance, starting from the $T'=5$\textsuperscript{th} frame. Note that the way in which ST-Chips are extracted could vary, and the use of optical flow in this prototype algorithm is just one of many ways in which these chips could be defined for the task of quality assessment. 

\begin{table*}
\caption{Descriptions of features in ChipQA.}
\begin{center}
\begin{tabular}{|l|l|l|}
\hline
Domain &  Description & Feature index\\
\hline
ST-Chip &  Shape and scale parameters from GGD fits at two scales. & $f_1-f_4$ \\
\hline
ST-Chip &  Four parameters from AGGD fitted to pairwise products at two scales. & $f_5-f_{36}$ \\
\hline
ST Gradient Chips &  Shape and scale parameters from GGD fits at two scales. & $f_{37}-f_{40}$ \\
\hline
ST Gradient Chips &  Four parameters from AGGD fitted to pairwise products at two scales. & $f_{41}-f_{72}$ \\
\hline
Spatial & Features and scores of spatial naturalness index NIQE. & $f_{73}-f_{109}$ \\
\hline
\end{tabular}
\label{feat}
\end{center}
\end{table*}

\section{Experiments and Results}

\subsection{Databases}
We evaluated our algorithm on four databases, as described in the following section.
\subsubsection{LIVE-APV Livestream VQA Database} This new database, which we call the LIVE-APV Livestream VQA database, and which will soon be released, was created for the purpose of developing tools for the quality assessment of live streamed videos. The database contains 367 videos, including both synthetically and authentically distorted videos. Of these, 52 videos are authentically distorted videos of different events. The remaining 315 videos were created by applying 6 different distortions on 45 unique contents. The 6 distortions are aliasing, compression, flicker, judder, interlacing, and frame drop. All the videos are of 4K resolution and were shown on a 4K TV in a human study in which 37 subjects participated.

\subsubsection{Konvid-1k} Konvid-1k\cite{konvid} is a database of 1200 videos of user-generated content with authentic distortions. Many videos in this database do not have significant temporal variation, and it has been found that NR IQA algorithms applied frame-by-frame often achieve high performance on Konvid-1k without making use of temporal information. All videos are of resolution 960$\times$540.

\subsubsection{LIVE Video Quality Challenge (VQC)} The LIVE VQC database\cite{vqc} has 585 videos of authentically distorted videos, each labeled by an average of 240 human opinion scores. The videos are of user-generated content.
\subsubsection{LIVE Mobile Database} The LIVE Mobile\cite{livemobile} database has 200 distorted videos created from 10 reference videos. The synthetically applied distortions are compression, wireless-packet loss, temporally varying compression levels, and frame-freezes. This study was conducted on mobile and tablet devices.

\subsection{Training Protocol}
The LIVE-APV database contains a mix of synthetically and authentically distorted content. Each algorithm was tested on 1000 random train-test splits, where 80\% of the data was used for training, and 20\% for testing. 5-fold cross validation was performed while ensuring content separation between training and validation data for each fold. Videos of the same content were not allowed to mix between folds. On the other databases, we implemented a 80-20 train-test split. A support vector regressor (SVR) was used to learn mappings from features to mean opinion scores. Cross validation was used to find the best parameters for the SVR for each algorithm. NIQE and VIIDEO are completely blind algorithms and were not trained, but were evaluated against the test set. 

We report the Spearman's rank ordered correlation coefficient (SROCC) and the Pearson’s linear correlation coefficient (LCC) between the scores predicted by the different algorithms and the subjective mean opinion scores. The predicted score was passed through the non-linearity described in\cite{logistic} before the LCC was computed. We report results for 1000 splits for all algorithms on the LIVE-APV database, and 100 splits on all other databases. Results are shown in Tables \ref{live}, \ref{konvid}, \ref{mobile}, and \ref{vqc}. We did not compute NR IQA features for every frame of each video in the databases, but computed them for at least 1 frame every second for each video. We averaged the features obtained by the NR IQA algorithms across frames and trained them with an SVR to map to mean opinion scores.

\begin{table}
\caption{Median SROCC and LCC for 1000 splits on the LIVE-APV Livestream VQA database}
\begin{center}
\begin{tabular}{|l|l|l|}
\hline
\textsc{Method}  &  SROCC & LCC \\
\hline
NIQE\cite{niqe} & 0.3395 & 0.4962 \\
\hline
BRISQUE\cite{brisque} (1 fps) &  0.6224 & 0.6843 \\
\hline
HIGRADE\cite{higrade} (1 fps) & 0.7159 & 0.7388  \\
\hline
CORNIA\cite{cornia} (1 fps) & 0.6778 & 0.7076  \\
\hline
TLVQM\cite{tlvqm} &  0.7597 & 0.7743\\
\hline
VIIDEO\cite{viideo} & -0.0039 & 0.2155\\
\hline
V-BLIINDS\cite{vbliinds} & 0.7264 & 0.7646 \\
\hhline{|=|=|=|} 
Spatial &  0.6770 & 0.7370 \\
\hline
ST-Chips & 0.6742 & 0.7235 \\
\hline
ST Gradient Chips & 0.7450 & 0.7611 \\
\hhline{|=|=|=|} 
\textbf{ChipQA-0} & \textbf{0.7802} & \textbf{0.8054}  \\
\hline
\end{tabular}
\label{live}
\end{center}
\end{table}

\begin{table}
\caption{Median SROCC and LCC for 100 splits on the Konvid database}
\begin{center}
\begin{tabular}{|l|l|}
\hline
\textsc{Method}  &  SROCC/LCC \\
\hline
NIQE\cite{niqe} & 0.3559/0.3860 \\
\hline
BRISQUE\cite{brisque} (1 fps) & 0.5876/0.5989 \\
\hline
HIGRADE\cite{higrade} (1 fps) & 0.7310/0.7390   \\
\hline
FRIQUEE\cite{friquee} (1 fps) & 0.7414/0.7486   \\
\hline
CORNIA\cite{cornia} (1 fps) & 0.7685/0.7671  \\
\hline
TLVQM\cite{tlvqm} &  \textbf{0.7749/0.7715} \\
\hline
VIIDEO\cite{viideo} & 0.3107/0.3269 \\
\hline
V-BLIINDS\cite{vbliinds} & 0.7127/0.7085 \\
\hline
ChipQA-0 & 0.6973/0.6943  \\
\hline
\end{tabular}
\label{konvid}
\end{center}
\end{table}

\begin{table}[ht]
\caption{Median SROCC and LCC for 100 splits on the LIVE Mobile database}
\begin{center}
\begin{tabular}{|l|l|}
\hline
\textsc{Method} &  SROCC/LCC \\
\hline
BRISQUE\cite{brisque} (1 fps) & 0.4876/0.5215 \\
\hline
VIIDEO\cite{viideo} & 0.2751/0.3439 \\
\hline
VBLIINDS\cite{vbliinds} & 0.7960/0.8585 \\
\hline
TLVQM\cite{tlvqm} & \textbf{0.8247/0.8744} \\
\hline
ChipQA-0 & 0.7898/0.8435  \\
\hline
\end{tabular}
\label{mobile}
\end{center}
\end{table}

\begin{table}
\centering
\caption{Median SROCC and LCC for 100 splits on the LIVE VQC database}\label{vqc}
\begin{tabular}{|l|l|}
\hline
\textsc{Method}  &  SROCC/LCC \\
\hline
BRISQUE\cite{brisque} (1 fps) & 0.6192/0.6519 \\
\hline
VIIDEO\cite{viideo} & -0.0336/-0.0064 \\
\hline
VBLIINDS\cite{vbliinds} & 0.7005/0.7251 \\
\hline
TLVQM\cite{tlvqm} & \textbf{0.8026/0.7999} \\
\hline
ChipQA-0 & 0.6692/0.6965  \\
\hline
\end{tabular}
\end{table}

\begin{table}
\centering
\caption{Computation time for a single 3840x2160 video with 210 frames from the LIVE-APV Livestream VQA database}\label{compcost}
\begin{tabular}{|l|l|}
\hline
\textsc{Method}  &  Time (s) \\
\hline
BRISQUE\cite{brisque} & 273 \\
\hline
HIGRADE\cite{higrade} & 14490 \\
\hline
CORNIA\cite{cornia} & 1797\\
\hline
FRIQUEE\cite{friquee} & 924000\\
\hline
VIIDEO\cite{viideo} & 4950 \\
\hline
VBLIINDS\cite{vbliinds} & 10774 \\
\hline
TLVQM\cite{tlvqm} & 892 \\
\hline
ChipQA-0 & 2284  \\
\hline
\end{tabular}
\end{table}
\subsection{Performance}
ChipQA-0 performs better than competing algorithms on the new LIVE-APV database, in which there are a number of commonly occurring temporal distortions, such as interlacing, judder, frame drop, and temporal variation of compression levels. We found the individual performance of each feature space  on the LIVE-APV database, and the results are shown in Table \ref{live}. It is clear that ST-Chips and ST Gradient chips provide  quality-aware information that boosts the performance of the algorithm on fast-moving, livestreamed videos. Studies have shown that UGC videos are dominated by spatial distortions\cite{liheng}. Both Konvid-1k and LIVE VQC are known to not contain  significant temporal variation and therefore NR IQA algorithms perform well on them without the need for temporal information or processing. Nevertheless, ChipQA-0 achieves competitive performance on these databases as well. It also performs competitively on the LIVE Mobile database, which has a mix of spatial and temporal distortions.

We perform a one-sided \textit{t} test on the 1000 SROCCs of the various algorithms on the live streaming database with a $95\%$ confidence level to evaluate the statistical signifiance of the results. The results show that ChipQA-0 is statistically superior to all other algorithms on the LIVE-APV Livestream VQA database.

\color{black}
\begin{table*}
\caption{Results of one-sided t-test performed between SROCC values of various algorithms on the LIVE-APV database. '1' ('-1') indicates that the row algorithm is statistically superior (inferior) to the column algorithm. The matrix is symmetric}
\begin{center}
\begin{tabular}{|l|l|l|l|l|l|l|l|l|}
\hline
\textsc{Method}  &  NIQE & BRISQUE & HIGRADE & CORNIA & TLVQM & VIIDEO &  V-BLIINDS & ChipQA-0  \\
\hline
NIQE & - & -1 & -1 & -1 & -1 & 1 & -1 & -1 \\
\hline
BRISQUE & 1 &  - & -1 & -1 & -1 & 1 & -1 & -1 \\
\hline
HIGRADE  &  1 & 1 & - & 1 & -1 & 1 & -1 & -1  \\
\hline
CORNIA  & 1 & 1 & -1 & - & -1 & 1 & -1 & -1  \\
\hline
TLVQM &  1 & 1 & 1 & 1 & - & 1 & 1 & -1\\
\hline
VIIDEO &  -1 &-1 & -1 & -1 & -1 & - & -1 & -1\\
\hline
V-BLIINDS & 1 & 1 & 1 & 1 & -1 & 1 & - & -1 \\
\hline
ChipQA-0 &  1 & 1 & 1 & 1 & 1 & 1 & 1 & -  \\
\hline
\end{tabular}
\label{stat}
\end{center}
\end{table*}

\color{black}
\subsection{Computational cost}
Table \ref{compcost} shows the computation times required to extract features for each algorithm on a single 4K video from the LIVE-APV database. Costs for the IQA algorithms were estimated by multiplying the computation time for a single frame by the total number of frames in the video. VIIDEO, VBLIINDS, and ChipQA-0 were implemented with Python. The other algorithms were implemented with MATLAB\textsuperscript{\textregistered}. It is not possible to directly compare these numbers because they were implemented on different platforms with different optimization strategies, but they can serve as a rough estimate, and ChipQA-0 is reasonably efficient. It was run on an Intel i9 9820X CPU with 10 cores and a maximum frequency of 4.1 GHz. All other algorithms were run on a AMD Ryzen 5 3600 with a maximum frequency of 4.2 GHz.

\color{black}
\section{Conclusion}
We have proposed the novel concept of ST-Chips, and defined how they are extracted and described why they are relevant to video quality. We used the statistics of these chips to model 'naturalness' and deviations from naturalness, and proposed parameterized statistical fits to their statistics. We further used the parameters from these statistical fits to map videos to subjective opinions of video quality without explicitly finding distortion-specific features and without reference videos. We showed that our prototype distortion-agnostic, no-reference video quality assessment algorithm, ChipQA-0, is highly competitive with other state-of-the-art models on a number of databases. We continue to refine the model, with one aim being to eliminate the need for an optical flow algorithm.

\section{Acknowledgments}

The authors thank the Texas Advanced Computing Center (TACC) at The University of Texas at Austin for providing HPC resources that have contributed to the research results reported in this paper. URL: http://www.tacc.utexas.edu.

\bibliographystyle{IEEEtran}
\bibliography{IEEEabrv,conference_101719.bib}

\end{document}